\documentclass[10pt,twocolumn,letterpaper]{article}

\usepackage[pagenumbers]{wacv} 

\usepackage{graphicx}
\usepackage{amsmath}
\usepackage{amssymb}
\usepackage{booktabs}
\usepackage{placeins}
\usepackage{stfloats}

\usepackage[pagebackref,breaklinks,colorlinks]{hyperref}

\usepackage[capitalize]{cleveref}
\crefname{section}{Sec.}{Secs.}
\Crefname{section}{Section}{Sections}
\Crefname{table}{Table}{Tables}
\crefname{table}{Tab.}{Tabs.}

\def\wacvPaperID{1760} 
\def\confName{WACV}
\def\confYear{2025}

\begin{document}
\title{Population-Scale Segmentation of Penile Tissue in DIXON MRI using Deep Learning for Quantitative Phenotyping in Male Reproductive Health}
\author{
Jan Ernsting$^{*}$\\
University of Münster\\
Germany\\
{\tt\small j.ernsting@uni-muenster.de}
\and
Gunnar Paul Kordes$^{*}$\\
University of Münster\\
Germany\and 
Nils Johannaber\\
University of Münster\\
Germany\and 
Lynn Ogoniak\\
University of Münster\\
Germany\and
Wolfgang Roll\\
University of Münster\\
Germany\and
Tim Hahn\\
University of Münster\\
Germany\and 
Alexander Siegfried Busch\\
University of Münster\\
Germany\and 
Benjamin Risse\\
University of Münster\\
Germany\\
{\tt\small b.risse@uni-muenster.de}
}
\maketitle

\renewcommand{\thefootnote}{}
\footnotetext{$^{*}$Equal contribution.}
\renewcommand{\thefootnote}{\arabic{footnote}}

%
%


\maketitle              
\begin{abstract}
Penile measurement is clinically relevant across male reproductive and urogenital health, including conditions such as micropenis, congenital and endocrine disorders, and sexual or urinary dysfunction.
However, quantitative assessment of penile size has relied mainly on external length or circumference measurements, which are difficult to standardize, sensitive to measurement conditions, and unable to capture the internal portion of the penis.
MRI enables volumetric assessment of the whole penis in vivo, but automated segmentation has not previously been established at population scale.
Automated whole-organ volumetry would enable high-throughput phenotyping for multi-omics and clinical studies of male reproductive disease.

Here, we present a deep learning framework for whole-penis segmentation in multi-channel DIXON MRI.
Using a newly curated expert-annotated training dataset ($n = 145$ subjects; $13,050$ annotated slices) and a double-annotated independent test benchmark ($n = 24$ subjects; $2,160$ double-annotated slices), we optimized a 3D nnU-Net architecture.
The model achieved a 5-fold cross-validation Dice score of $0.90$ and performed at observer-level accuracy on the independent test set (Dice: $0.92$; Hausdorff distance: $3.58$).

We deployed the model in $34,412$ UK Biobank participants, enabling automated quantification of total penile tissue, including both external and internal components. 
Longitudinal evaluation in 2,282 men demonstrated high inter-session reproducibility ($r = 0.87$). 
This framework establishes a reproducible and population-scalable method for MRI-based assessment of penile anatomy and provides an open technical resource for future studies in urological imaging and male reproductive health. 
The trained model weights will be publicly released.


\end{abstract}
\section{Introduction}
Male genital anatomy exhibits substantial inter-individual variation in shape and size, which may be influenced by developmental, hormonal, metabolic, and aging-related processes.
Quantitative characterization is therefore relevant for biomedical research and clinical medicine, supporting the the assessment of anatomical variation and abnormalities. 
However, standardized large-scale assessment of penile morphology remains limited.
Existing population-level evidence relies predominantly on observer-dependent external measurements ~\cite{veale2015am-abe} that do not capture the proximal penile root, comprising the crura and bulb, despite its anatomical and clinical relevance to total erectile tissue~\cite{kumar2025mr-a44}.
This limitation is especially relevant for conditions affecting penile development or presentation, such as micropenis, buried penis, and androgen insensitivity syndrome, where whole-organ morphology may provide information beyond external dimensions alone~\cite{hiort2014management-95a}.

Magnetic resonance imaging (MRI) offers the possibility of non-invasive three-dimensional characterization of penile anatomy, enabling visualization of both external and internal components. 
In particular, large population imaging resources such as the UK Biobank provide standardized Dixon MRI acquisitions that create unprecedented opportunities for systematic morphometric analysis across thousands of participants.
However, accurate penis segmentation is complicated by the distinct challenges of the external and proximal penile compartments:
the pendulous penis shows substantial variation in shape and position across scans, whereas the penile root is anatomically embedded within the perineum and closely bordered by adjacent muscles and pelvic soft tissues.
These properties make whole-organ delineation more complex than segmentation of many spatially constrained internal organs.

Deep learning has become a key methodology for medical image segmentation and has enabled large-scale quantitative phenotyping in biomedical imaging.
While recent frameworks have successfully segmented thoracic, abdominal, and pelvic structures \cite{graf2025vibesegmentator-167}, automated segmentation of penile tissue remains unexplored.
To address this gap, we introduce a deep learning-based framework for population-scale penis segmentation in UK Biobank Dixon MRI.


\subsection{Contribution}
The primary contributions of this work are threefold:
First, we introduce a novel, manually annotated dataset for penile MRI segmentation derived from the large-scale UK Biobank cohort. 
Second, we provide a highly reliable evaluation benchmark consisting of an independent test set independently annotated and verified by two domain experts, ensuring a robust ground truth for future comparative studies. 
Third, to facilitate reproducibility and accelerate future research in automated pelvic imaging, we publicly release our finalized deep learning model configurations and trained weights.

Together, these contributions enable population-scale multi-omics and clinical association studies of whole-penis volume, providing a scalable imaging phenotype for translational research in male reproductive and urogenital disease.

\section{Related Works}

\subsection{Segmentation in MRI data}

Deep learning has fundamentally transformed semantic segmentation in medical imaging. 
Foremost among segmentation frameworks suited for medical data is the nnU-Net framework~\cite{isensee2021nnu-net-890}, which has established state-of-the-art benchmarks by automating data preprocessing, network topology configuration, and training hyperparameters based on dataset properties.

A notable advancement in this domain is VIBESegmentator~\cite{graf2025vibesegmentator-167}, which was developed to address automated segmentation in volumetric Dixon MRI used in large-scale cohorts such as NAKO and UK Biobank for full body MRI segmentation.
Tools like VIBESegmentor demonstrate the feasibility of processing high-throughput, multi-station abdominal and pelvic datasets to extract reliable morphological segmentations.

However, despite these advances in overall body composition and tissue segmentation, highly localized structures within the male pelvic region remain underrepresented. 
Addressing this gap, Ernsting et al.~\cite{ernsting2025towards-8ff} recently demonstrated the utility of targeted deep learning models for male reproductive anatomy, specifically focusing on testis segmentation within the UK Biobank. 
While their work demonstrates usage of Dixon MRI data to segment small pelvic structures, the segmentation of penile anatomy remains an unaddressed challenge due to high structural variability, complex surrounding boundaries, and a lack of expert-annotated training data. 
This study directly extends previous efforts by introducing a dedicated framework and expert-verified benchmark for automated penis segmentation.

\subsection{Evaluation of penile size and morphology}


In clinical routine, assessment of penile size is performed for several distinct indications, including the evaluation of differences or disorders of sex development, congenital genital anomalies such as hypospadias or buried penis, endocrine disorders associated with impaired androgen production or action, the diagnosis of micropenis, counselling of men with perceived penile shortness, evaluation of erectile and acquired structural abnormalities, and surgical planning and follow-up.
Conventional methods primarily rely on anthropometric measurements, including flaccid penile length, stretched penile length, erect length, and circumference~\cite{veale2015am-abe}. 
However, these clinical metrics remain substantially method-dependent, as measurements vary according to anatomical landmarks, stretching force, and observer experience, introducing considerable inter-observer variability and limiting comparability across studies~\cite{101016jjsxm202011012}.
Furthermore, conventional measurements are restricted to external dimensions and do not capture the complete anatomical and volumetric extent of the penis. 
Ultrasonography provides additional structural information but is mainly applied for selected anatomical or hemodynamic assessments and lacks a standardized framework for whole-organ quantification.
MRI, although not routinely used for penile size evaluation, enables volumetric visualization of the complete penile anatomy, including both external and internal structures such as the glans, corpora cavernosa, corpus spongiosum with bulb, and crura~\cite{kumar2025mr-a44}. 
Consequently, MRI-based AI segmentation represents a promising approach for standardized, observer-independent quantification of total penile tissue volume and morphology.
While AI-assisted contouring of individual penile structures has shown feasibility in MRI-based studies~\cite{berg2023deep-85b,schubert2025advancing-b78}, 
comprehensive penis segmentation remains largely unexplored.

\section{Methods}

\subsection{Dataset}

%

Data for this study were obtained from the UK Biobank, a large-scale, prospective biomedical database and research resource. 
Eligibility for the current analysis was restricted to male participants who completed the abdominal MRI imaging protocol from 2014 onward. 
To ensure data integrity, the cohort was filtered to include only individuals with concordant self-reported and genetically inferred sex. 
This yielded a primary study population of $34,600$ male participants aged $45-87$ years (mean: $67$, standard deviation [SD]: $8$). 
Within this cohort, a subset of $2,315$ participants underwent repeat imaging assessments from 2019 onward ($1 - 10$ years time interval; mean: $5$ years; SD: $2$ years), providing a longitudinal follow-up subcohort.
After preprocessing, a final sample of $34,412$ participants was utilized for the penis segmentation task.
For model training, only the initial imaging sessions were considered.

The Dixon MRI data from the UK Biobank capture complementary anatomical structures across six spatial imaging slabs extending from the neck to the knee, and covering a total height of 1.10 meters~\cite{west2016feasibility-b44}.
This MRI technique generates four distinct volumetric reconstructions per slab: water, fat, in-phase, and out-of-phase images.
For image processing and analysis, the original DICOM files were converted into the NIfTI format.


\subsection{Preprocessing}
Since the relevant anatomy spanned multiple imaging slabs, a preprocessing pipeline was implemented to combine the relevant acquisitions. 
Specifically, the two lowermost caudal imaging slabs of all four sequence modalities (water, fat, in-phase, and opposed-phase) were stitched together to form a continuous volumetric field of view.
Stitching was performed using an open-source rigid registration framework~\cite{lavdas2019machine-974}. 
The registration pipeline was configured with a 2-pixel margin ignored at the image edges during alignment and voxel intensity averaging enabled across the overlapping regions to ensure smooth transitions between the stations and to remove artifacts.

\subsection{Annotation}



Manual annotations were performed using 3D Slicer (version 5.10.0; \url{https://www.slicer.org})~\cite{fedorov20123d-c59} on opposed-phase sequences, ensuring that the penis was completely visible within the field of view across all included scans. 
The total dataset consisting of $169$ randomly selected images from the UK Biobank dataset was partitioned into a training set ($n=145$) and a test set ($n=24$).
The training set was divided between two independent readers, where Annotator 1 segmented $108$ cases ($9,720$ annotated axial slices) and Annotator 2 segmented $37$ cases ($3,330$ annotated axial slices). 
Conversely, the test set consisted of $24$ MRIs ($2,160$ double-annotated axial slices) dedicated to double annotation, where both Annotator 1 and Annotator 2 independently segmented the same cases to facilitate rigorous inter-observer validation.
These annotations were used to measure inter-annotator agreement and define the baseline for human annotation performance.

To resolve discrepancies and break ties between these double segmentations for final model evaluation, a third imaging expert blindly reviewed and selected the superior annotation for each case.
This final set of $24$ images is the test set used for model evaluation.



\begin{figure*}
    \centering
    \includegraphics[width=.9\textwidth]{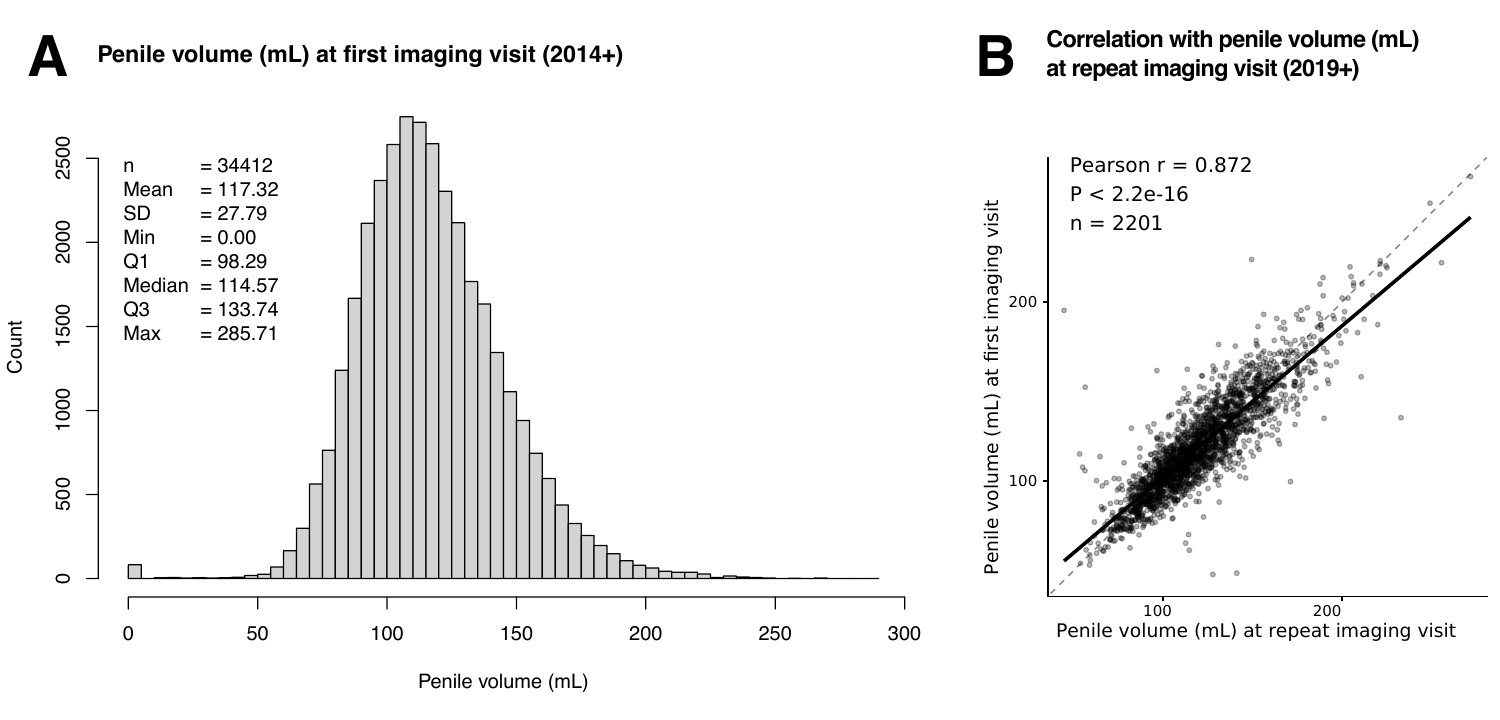}
    \caption{(A) Predicted volume in milliliters (mL) for $n=34,412$ UK Biobank subjects.
    Predictions with 0 mL were manually reviewed and identified as imaging errors, i.e. the penis not being visible in the merged slabs.
    (B) Correlation between predicted volumes at two measurement sessions excluding subjects with zero-volume prediction. For $n=2,201$ subjects a second imaging session was performed by UK Biobank (mean interval was $5$ years, SD = $2$). 
    Correlation between predicted volume at first and repeat imaging session is $r=0.87$} 
    \label{fig_results_plots}
\end{figure*}

\subsection{Model}
For the segmentation of penile structures, we employed the nnU-Net framework~\cite{isensee2021nnu-net-890}, adapting the methodological pipeline proposed by Ernsting et al.~\cite{ernsting2025towards-8ff}.
The nnU-Net framework is a self-configuring meta-architecture that automatically tailors preprocessing, network topology, and training hyperparameters to the specific dataset characteristics, eliminating manual heuristic tuning and ensuring highly reproducible segmentation pipelines.

We trained and compared both the 2D and 3D full-resolution configurations of the architecture.
The network architectures were configured to accept a four-channel input consisting of in-phase, opposed-phase, water, and fat images, which were mapped to the corresponding target binary segmentation masks.

Training was conducted strictly adhering to the default nnU-Net configuration.

\subsection{Training}
Model training and inference were executed on a computing system equipped with an AMD EPYC 9124 CPU and an NVIDIA GeForce RTX 4090 GPU (24 GB VRAM). 

The experimental pipeline was structured into three phases:
First, internal model training and architectural tuning were conducted using a 5-fold cross-validation scheme on the training dataset. 
Second, to evaluate generalization performance, model inference was executed on the test dataset.
Third, the model was deployed to perform inference across the $34,600$-participant cohort, generating the segmentation masks for subsequent large-scale population analysis.

\section{Results}

\subsection{Human performance}

To establish a baseline for human performance and evaluate inter-observer variability, the $24$ holdout images in the test dataset were independently segmented by two domain experts. 
The spatial overlap between the two manual annotations yielded a mean Dice similarity coefficient (DSC) of $0.82$ SD$=0.02$ (median: $0.82$).
To evaluate boundary distances, the Hausdorff Distance was calculated, revealing a mean boundary separation of $9.54$ SD$=3.54$~mm (median: $9.38$~mm). 

\begin{figure}
    \centering
    \includegraphics[width=0.4\textwidth]{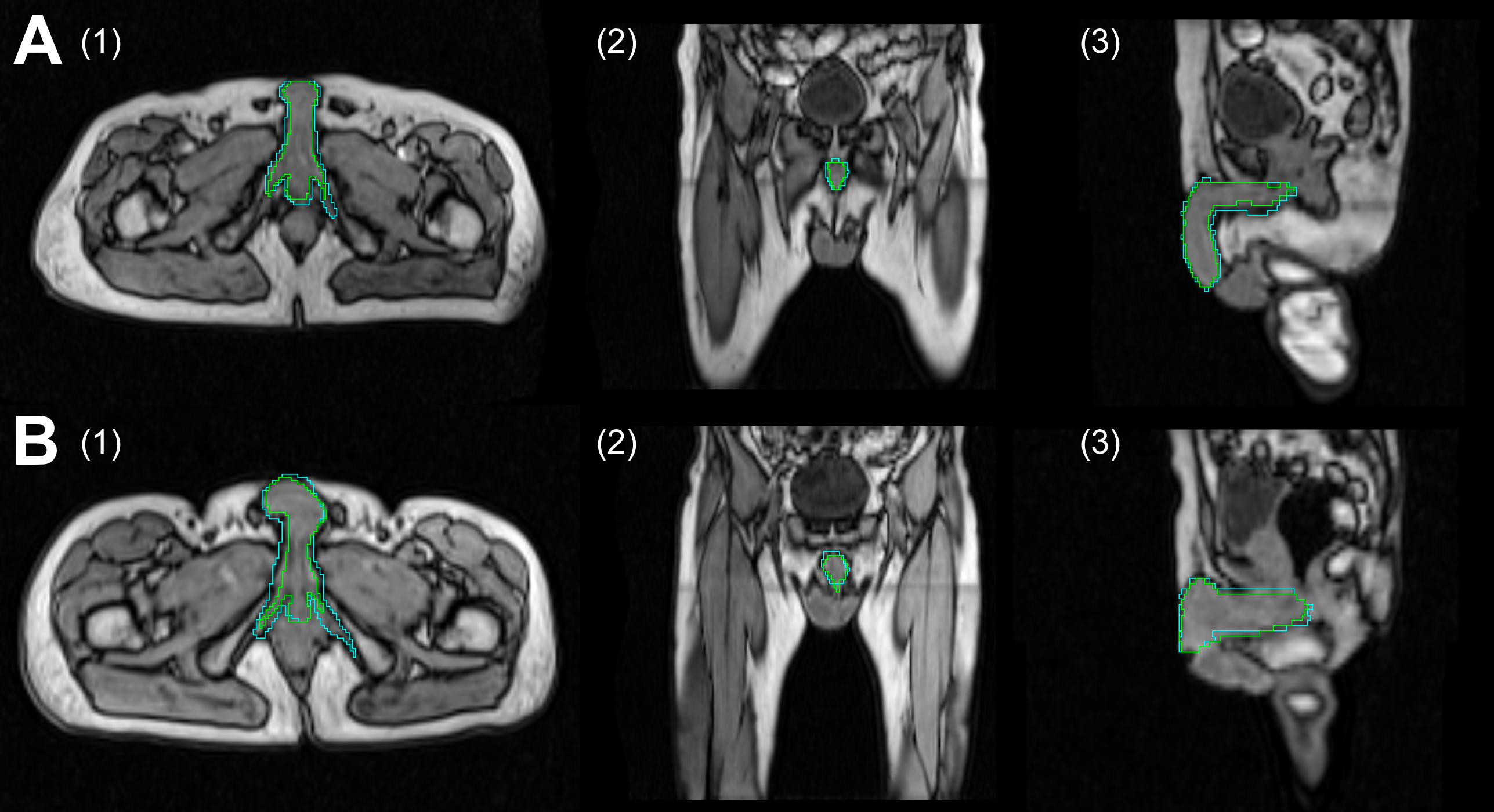}
    \caption{Example annotations of penile tissue in UK Biobank DIXON MRI opposed-phase acquisitions using merged imaging slabs.
    Axial (1), coronal (2), and sagittal (3) views are shown.
    Green annotations indicate Annotator 1, and blue annotations indicate Annotator 2.
    Inter-annotator differences are primarily located in the proximal penile root, including the crura and bulb.
    Images reproduced by kind permission of UK Biobank \textcopyright.} 
    \label{fig_annotations}
\end{figure}

\subsection{5-Fold Cross-Validation results}
Both 2D and 3D network configurations were evaluated to determine the optimal dimensionality for the segmentation task. 

The 3D nnU-Net configuration demonstrated superior performance and stability across all folds, achieving a mean DSC of $0.90$ SD$=0.002$ (median: $0.90$).
In comparison, the 2D configuration yielded a slightly lower but competitive spatial overlap, with a mean DSC of $0.89$ SD$=0.005$ (median: $0.89$).
Notably, the remarkably low standard deviation observed in the 3D ($0.002$) and 2D model ($0.005$) highlights the consistency across the folds. 
Based on these cross-validation outcomes, the 3D model was selected as the primary architecture for subsequent test dataset benchmarking and population-scale deployment.



\subsection{Test performance}
The network predictions were benchmarked against the ground truth segmentations on the 24 test images where the imaging specialist broke ties. 
The model achieved a high spatial overlap, yielding a mean DSC of $0.92$ SD$=0.03$ (median: $0.93$). 
Boundary distance evaluations via the Hausdorff Distance demonstrated tight contour alignment, with a mean boundary separation of $3.58$ SD$=2.44$ mm (median: $2.72$ mm). 
The mean absolute volume difference between the manual and predicted segmentations was $8249.09$ SD$=9289.66$ voxels (median: $5739.80$ voxels). 
\begin{figure*}[!t]
    \centering
    \includegraphics[width=.9\textwidth]{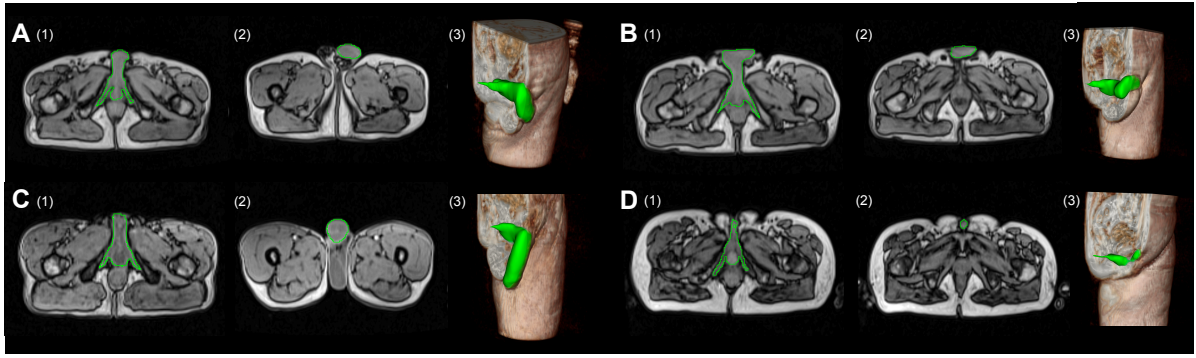}
    \caption{Examples of nnU-Net predictions. Axial views highlighting the internal (1) and external penile compartments (2), along with corresponding 3D renderings (3), are shown. (B) illustrates likely compression of the external penis due to clothing. (D) depicts a sample with visible internal erectile structures but no externally developed penile shaft. Images reproduced by kind permission of UK Biobank \textcopyright.} 
    \label{fig_predicted_volumes}
\end{figure*}
\subsection{Population scale results}
To demonstrate the scalability and robustness of the developed segmentation network, inference was done at a population scale on the entire cohort of $34,412$ subjects from the UK Biobank dataset.
Figure~\ref{fig_predicted_volumes} shows sample model predictions on unseen data.
Figure~\ref{fig_results_plots}~A illustrates the distribution of the predicted anatomical volumes in milliliters (mL).
During initial processing, a subset of cases yielded predicted volumes of zero. 
A manual review of these statistical outliers revealed that they exclusively represented artifactual imaging errors rather than algorithmic failures.
Specifically, these zero-volume predictions corresponded to cases where the target anatomy was completely omitted from the field of view due to the positioning of the imaging station or misalignment during the merging of the acquisition slabs.




To evaluate the longitudinal consistency of the network's predictions, we analyzed a sub-cohort of $n=2,282$ subjects who underwent a second, independent imaging session as part of the UK Biobank longitudinal assessment protocol.
After excluding participants with zero-volume segmentations in either session ($n=81$), $n=2,201$ subjects were retained.
Volumetric predictions from the two sessions showed high agreement ($r=0.87$; Figure~\ref{fig_results_plots}~B), indicating good test-retest reliability of the model across independent acquisitions.

\section{Discussion}





This work demonstrates that penile measurements can be converted from a difficult-to-standardize observation into a reproducible, population-scale image-derived phenotype.
Its central contribution is the establishment of a volumetric segmentation model capturing both external and internal penile components which are entirely inaccessible to conventional surface anthropometry.
The model exceeded inter-observer agreement on the independent test set (Dice: $0.92$ vs. $0.82$).
The relatively large inter-observer gap reflects genuine anatomical ambiguity at the proximal root, where expert boundaries tend to diverge (Figure \ref{fig_annotations}).
Zero-volume predictions in population deployment were attributable to field-of-view errors, not model failures.
Several limitations apply.
The training set is modest and the cohort skews older (mean 67 years), which may limit generalizability.
The model segments the penis as a single structure; substructure-level delineation (glans, corpora, bulb, crura) would be valuable but is constrained by Dixon MRI resolution.
Future work should validate in younger and clinical cohorts and explore sub-organ segmentation as resolution improves.
The public release of model weights enables replication and extension across independent cohorts, supporting large-scale association studies linking penile morphology to genetics, hormonal profiles, body composition, and reproductive health.

\textbf{Prospects of Application: }
This framework introduces penile volume as a scalable image-derived phenotype for population-based male health research. It enables large-scale association studies linking penile anatomy to genetic variants, proteomic and metabolomic profiles, hormonal status, body composition, aging, and reproductive or urogenital traits, thereby advancing research into the biological basis of male reproductive disorders.

    

\section{Acknowledgments}
J.E. was supported by the Medical Scientist Kolleg InFlame funded by the Else Kröner-Fresenius Foundation, Germany.

ASB is funded by the German Research Foundation (Deutsche Forschungsgemeinschaft, DFG) – 464240267 and SFB 1748/1 2026.

This research has been conducted using the UK Biobank Resource under Application Number 100261.

%
{\small
\bibliographystyle{ieee_fullname}
\bibliography{references}
}
\end{document}